\documentclass[12pt,preprint]{aastex}

\slugcomment{accepted by ApJ Letters}

\begin{document}

\title{Molecular Gas and Star Formation in the Cartwheel}

\author{James L. Higdon\altaffilmark{1}, Sarah J. U. Higdon\altaffilmark{1},
            Sergio Mart\'in Ruiz\altaffilmark{2}, and Richard J. Rand\altaffilmark{3} }

\altaffiltext{1}{Department of Physics, Georgia Southern University, Statesboro, GA 30460-8031, USA}
\altaffiltext{2}{Joint ALMA Office, Alonso de Co\'rdova 3107, Vitacura, Casilla 19001,
                 Santiogo 19, Chile}
\altaffiltext{3}{Department of Physics and Astronomy, University of New Mexico, Albuquerque, NM 87131, USA}

\begin{abstract}
Atacama Large Millimeter/submillimeter Array (ALMA) $^{12}$CO(J=1-0) observations 
are used to study the cold molecular ISM of the Cartwheel ring galaxy 
and its relation to HI and  massive star formation (SF). CO moment maps find  
$(2.69\pm0.05)\times10^{9}$  M$_{\odot}$ of H$_2$ associated with the 
inner ring ($72\%$) and nucleus ($28\%$) for a Galactic $I_{\rm CO}$-to-$N_{\rm H_2}$ 
conversion factor ($\alpha_{\rm CO}$). The spokes and disk are not detected.
Analysis of the inner ring's CO kinematics show it to be expanding
($V_{\rm exp}=68.9\pm4.9$ km s$^{-1}$) implying an $\approx70$ Myr age.
Stack averaging reveals CO emission in the starburst outer ring for the first time,
but only where HI surface density ($\Sigma_{\rm HI}$) is high, representing 
$M_{\rm H_2}=(7.5\pm0.8)\times10^{8}$ M$_{\odot}$ for a metallicity appropriate 
$\alpha_{\rm CO}$, giving small $\Sigma_{\rm H_2}$ ($3.7$ M$_{\odot}$ pc$^{-2}$), 
molecular fraction ($f_{\rm mol}=0.10$), and H$_2$ depletion 
timescales ($\tau_{\rm mol} \approx50-600$ Myr). Elsewhere in the outer ring
$\Sigma_{\rm H_2}\lesssim 2$ M$_{\odot}$ pc$^{-2}$, $f_{\rm mol}\lesssim 0.1$ and 
$\tau_{\rm mol}\lesssim 140-540$ Myr (all $3\sigma$). The inner ring and nucleus are 
H$_2$-dominated and are consistent with local spiral SF laws. $\Sigma_{\rm SFR}$ 
in the outer ring appears independent of $\Sigma_{\rm H_2}$, $\Sigma_{\rm HI}$ 
or $\Sigma_{\rm HI+H_2}$.  The ISM's long confinement in the robustly star forming
rings of the Cartwheel and AM0644-741 may result in either a large diffuse H$_2$ component or 
an abundance of CO-faint low column density molecular clouds.  The H$_2$ content of evolved 
starburst rings may therefore be substantially larger. 
Due to its lower $\Sigma_{\rm SFR}$ and age the Cartwheel's inner ring has yet to 
reach this state. Alternately, the outer ring may trigger efficient SF in an HI-dominated ISM.
\end{abstract}

\keywords{galaxies: individual(Cartwheel) --- galaxies: interactions --- galaxies: ISM --- galaxies: starburst } 

\section{Introduction}

Gravitational interactions play a fundamental role in galaxy evolution, from their assembly at high-$z$
\citep[][]{whiterees78, jogee09} to much of their subsequent chemical and luminosity evolution
\citep[][]{smail97, tacconi08}. Detailed studies of local interacting galaxies can provide important 
insights into these processes, particularly if the dynamical history can be reconstructed
to allow detailed modeling. Ring galaxies, created in the passage of a companion through a spiral's disk, 
provide excellent examples. The interaction generates one or more outwardly traveling orbit-crowded rings 
that ``snow-plow'' the disk's ISM as they propagate \citep[][]{lt77, struckhigdon, mapelli2008} and
trigger star formation (SF) via large-scale gravitational instabilities \citep[][hereafter HHR]{wof2,wof4}.
The rings host high and sustained SF throughout their $100-400$ Myr lifetimes as evidenced by
numerical studies (see Fig.$~12$ in \citet[][]{mihos_hernquist} and Fig.$~9$ in 
\citet[][]{mapelli_mayer2012}) and observations of large optical/near-infrared
radial color gradients from aging post-starburst clusters left in the ring's wake
\citep[e.g.,][]{marcum93,appleton1996,romano2008}.

The ISM's long confinement in the rings provides a unique environment for 
studying the effects of massive stars on its state.
AM0644-741's starburst ring, for example, appears to possess a remarkably 
low $M_{\rm H_2}$, molecular fraction ($f_{\rm mol}=M_{\rm H_2}/(M_{\rm HI}+M_{\rm H_2})$) and 
mean gas volume density ($n_{\rm H}$) despite conditions favoring an H$_2$-dominated ISM, along
with a peculiar SF law in which $\Sigma_{\rm HI}$ but not $\Sigma_{\rm H_2}$ is correlated with 
$\Sigma_{\rm SFR}$ (HHR), i.e., a complete reversal of the situation in ordinary
spirals, e.g., Kennicutt et al. (2007; K07), Bigiel et al. (2008; B08), Schruba et al.
(2011; S11), and Leroy et al. (2013; L13). Is AM0644-741 unique 
or is its ISM characteristic of other large and robustly star forming ring galaxies? 
In this {\em Letter} we investigate the Cartwheel (Fig.~$1$), a similarly large ($45$ kpc), 
evolved (age $\approx R_{\rm ring}/V_{\rm exp} = 440$ Myr), and 
gas rich (M$_{\rm HI}=2.9\times10^{10}$ M$_{\odot}$) ring galaxy with comparable SFR 
\citep[][]{wof1,wof2} and - unique among ring galaxies - a small second ring. We combine
deep $^{12}$CO(J=1-0) observations with the Atacama Large Millimeter/submillimeter Array 
(ALMA) with optical, infrared and radio data from other telescopes. The molecular 
content of this galaxy has been a long-standing puzzle \citep[][]{horellou95,horellou,wof2}. 
ALMA's revolutionary capabilities make it possible to examine the Cartwheel's molecular 
and atomic ISM and young stellar populations on comparable physical scales, allowing direct 
comparisons with local spirals.

\section{Observations and Analysis}
The Cartwheel was mosaiced with ALMA using $7$-array pointings on $2013$ December $24$ 
(2012.1.00720.S) for uniform sensitivity to $^{12}$CO(J=1-0) emission.
The correlator provided a $468.8$~MHz ($1256$ km s$^{-1}$) total bandwidth  
with $244.2$ kHz ($0.65$ km s$^{-1}$) channels centered on the redshifted $115.2712$ GHz line. 
Antenna gains and bandpass were calibrated using J0026-3512 while observations
of Uranus set the absolute flux scale.
CASA~$4.2.1$ was used for reduction \citep[][]{mcmullin}.  A {\it natural}-weighted image cube 
consisting of $600$ channels separated by $2.0$ km s$^{-1}$ was created and deconvolved using 
{\em imager}, giving a $2.43\arcsec\times1.50\arcsec$ ($1.47\times0.91$ kpc) synthesized 
beam and a $1.39$ mJy beam$^{-1}$ rms. Integrated line flux (moment-$0$) and 
flux-weighted radial velocity (moment-$1$) maps are shown in Figs.$~1 ~\& ~2$. 
$M_{\rm H_2}$ is derived by multiplying the CO line luminosity ($L^{'}_{\rm CO}$)
\begin{equation}
{\rm
L'_{CO}=\left({{c^2}\over{2k_{\rm B}}}\right) \left({{S_{\rm CO}\Delta v}\over{Jy~km~s^{-1}}}\right) 
\nu_{\rm obs}^{-2} D^{2}_{\rm L} (1+z)^{-3}
}
\end{equation}
\citep[][]{solomonvdb05} by the appropriate $I_{\rm CO}$-to-$N_{\rm H_2}$ 
conversion factor: $\alpha_{\rm CO}$. All surface densities and masses are scaled by 
$1.36$ to account for helium.

We also utilize long-slit optical spectra of both rings from the Mt. Stromlo \& Siding 
Springs Observatory $2.3$m telescope's double-beam spectrograph, 
a calibrated H$\alpha$ emission-line image from the Canada France Hawaii Telescope, 
a MIPS $24~\mu$m image, and Very Large Array integrated HI and velocity maps \citep[][]{wof2}.

H$\alpha$ and $24~\mu$m emission is combined to derive internal extinction using
\begin{equation}
{\rm
A_{\rm H\alpha} = 2.5~log\left( 1~+~0.038{{\nu L_{\nu}(\rm 24~\mu m)}\over{L_{\rm H\alpha}}}   ~\right)
}
\end{equation}
(K07), where $L_{\nu}(24~\mu m)$ and $L_{\rm H\alpha}$ are measured within identical $12\arcsec$
apertures. We find $A_{\rm H\alpha} = 0.4-1.4$ around the outer ring (consistent with 
Fosbury \& Hawarden (1977)'s optical spectroscopy) and $1.9\pm0.3$ and $2.58\pm0.2$ for the inner ring and nucleus. 

Measured ${\rm {{[N~II]_{6584}}\over{H\alpha}}}$ and ${\rm {{[S~II]_{6716+6731}}\over{H\alpha}}}$ 
ratios are used to determine metallicity following \citet[][]{nagao} and constrain the
metallicity dependent $\alpha_{\rm CO}$. We find 12 + log(O/H) $=9.1$ in the inner ring
and $8.2-8.3$ for outer ring HII complexes with uncertainties 
of $\approx0.15$. Fig.~$3$ in \citet[][]{magdis} gives $\alpha_{\rm CO} = 25$ M$_{\odot}$  
(K km s$^{-1}$ pc$^{2}$)$^{-1}$ for the outer ring with a 
substantial spread reflecting the dispersion in individual measurements plus their 
uncertainties.  Consequently, the correct $\alpha_{\rm CO}$ may lie between $3.7-90.0$ 
M$_{\odot}$ (K km s$^{-1}$ pc$^{2}$)$^{-1}$. We adopt $\alpha_{\rm CO} = 25$ M$_{\odot}$  
(K km s$^{-1}$ pc$^{2}$)$^{-1}$ to derive $M_{\rm H_2}$, $\Sigma_{\rm H_2}$, and $f_{\rm mol}$ 
but give the full range implied by the $\alpha_{\rm CO}$ spread within parentheses.

\section{Results and Discussion}

\subsection{CO Moment Maps}
$115~$GHz line emission is directly detected only in the inner ring and nucleus (Fig.~$1$). 
We measure an integrated line flux of $12.86\pm0.28$ Jy km s$^{-1}$ in agreement with 
single dish observations \citep[][]{horellou}, implying that ALMA recovers essentially 
all of the Cartwheel's CO emission. $72\%$  originates in the inner ring where the 
clumpy gas distribution coincides with prominent dust lanes and luminous 
($L_{\rm H\alpha} \approx1-3\times10^{39}$ erg s$^{-1}$; A$_{\rm H\alpha}=1.9$) 
HII complexes. CO peaks are typically associated with H$\alpha$ sources. Given the
inner ring's metallicity and modest $\Sigma_{\rm SFR}$ we adopt a Galactic 
$\alpha_{\rm CO}$  \citep[$3.68$ M$_{\odot}$ (K km s$^{-1}$ pc$^{2}$)$^{-1}$;][]{strong}, giving 
a total $M_{\rm H_2}$ of $(1.94\pm0.04)\times10^{9}$ M$_{\odot}$ and 
$\Sigma_{\rm H_2}= 8 - 148$ M$_{\odot}$ pc$^{-2}$. The remaining H$_2$, $(7.5\pm0.1)\times10^{8}$ 
M$_{\odot}$ for a Galactic $\alpha_{\rm CO}$,
originates in an unresolved nuclear source representing the galaxy's highest
$\Sigma_{\rm H_2}$ ($\ge 216$ M$_{\odot}$ pc$^{-2}$). The 
inner ring and nucleus are both H$_2$-dominated ($f_{\rm mol}\approx1$).
CO emission is not evident anywhere else in the moment-$0$ map, implying
$\Sigma_{\rm H_2} \lesssim 4$ M$_{\odot}$ pc$^{-2}$ ($3\sigma$) for a Galactic $\alpha_{\rm CO}$. 

The inner ring's CO velocity field (Fig.$~2$) displays ordered rotational motion along with
streaming visible as kinks in the isovelocity contours. Although we expect the inner ring to 
be expanding we fit its radial velocity distribution with models
of rotating circular rings both with and without expansion ($V_{\rm exp}$;
see HHR). An {\em expanding}  circular ring ($i=56.6^{\circ}$) provides the best 
fit, giving $V_{\rm sys}=9138.5\pm2.7$, $V_{\rm circ}=170.3\pm5.2$ and $V_{\rm exp}=68.9\pm4.9$
km s$^{-1}$. For the inner ring's $4.8$ kpc radius an $\approx70$ Myr age is implied, 
or $\approx1/6^{\rm th}$ the outer ring's.

\subsection{CO Stacking Analysis}
Stack averaging is used to pursue faint $^{12}$CO(J=1-0) emission in the outer ring.
We first divide it into three annular sections (I, II $\&$ III in Fig.~$1$) of approximately 
equal projected area using the HI and H$\alpha$ images to define their width.  
Section II possesses both the Cartwheel's highest $\Sigma_{\rm HI}$ and $\approx50\%$ of
its total $M_{\rm HI}$.  CO spectra within a given section are coadded after being first 
shifted using the HI velocity at that position to place the $115$ GHz line at the origin. 
The averaged spectra are then smoothed  ($24$ km s$^{-1}$ boxcar) and rebinned ($8$ km s$^{-1}$). 
We report the first detection of molecular gas in the Cartwheel's outer ring in section II (Fig.~$3$)
and derive a mean CO intensity ($I_{\rm CO}$) of $1.40\pm0.16$ mJy km s$^{-1}$ arcsec$^{-2}$
and a noticeably asymmetric profile ($\Delta V_{\rm FWZI}=48$ km s$^{-1}$).  For our adopted 
$\alpha_{\rm CO}$ this represents $M_{\rm H_2} = (7.5\pm0.8)\times10^{8}$ M$_{\odot}$ 
($(0.8-25.3)\times10^{8}$ M$_{\odot}$) and a mean $\Sigma_{\rm H_2}$ of $3.7\pm0.4$ 
M$_{\odot}$ pc$^{-2}$ ($0.4-12.6$ M$_{\odot}$ pc$^{-2}$). In this section 
$M_{\rm HI} = 7.0\times10^{9}$ M$_{\odot}$, giving $f_{\rm mol} = 0.10\pm0.01$
($0.01-0.27$). CO emission is not detected from sections I or III after stacking 
($I_{\rm CO}$ $<1.00$ and $<0.71$ mJy km s$^{-1}$ arcsec$^{-2}$, respectively), implying
$M_{\rm H_2}\lesssim3\times10^{8}$ M$_{\odot}$ ($0.4-9.3\times10^{8}$ M$_{\odot}$),
$\Sigma_{\rm H_2}\lesssim 2-3$ M$_{\odot}$ pc$^{-2}$ ($0.3-10.8$ M$_{\odot}$ pc$^{-2}$) and
$f_{\rm mol}\lesssim0.08$ ($0.01-0.28$) on average over $\approx2/3$ of the $45$ kpc diameter outer ring.

To better define the H$_2$ distribution we subdivide the outer ring into eight annular 
sections ($a-h$ in Fig.~$4$) of $\approx100$ kpc$^{2}$ projected area
using the same widths as before. 
Two sections of special interest due to their intense SF - the ``northern arc'' and ``southern
quadrant'' \citep[$\approx80\%$ of $L^{\rm total}_{\rm H\alpha}$;][]{wof1} - are represented
by sections $a$ and $d+e$ respectively. Stack averaging is done
as before but with slightly increased smoothing ($30$ km s$^{-1}$ boxcar) and rebinning 
($10$ km s$^{-1}$). We detect $^{12}$CO(J=1-0) only in $b$, $c$ \& $d$. All three sections are 
characterized by high $\Sigma_{\rm HI}$ but a wide range in 
$\Sigma_{\rm SFR}$.  Emission is not detected in the northern arc and, unexpectedly, 
in only half of the  southern quadrant ($d$ but not $e$) despite similar $\Sigma_{\rm SFR}$
and $\Sigma_{\rm HI}$. We derive $I_{\rm CO}$ of $(0.54\pm0.11)$, 
$(1.28\pm0.24)$, and $(1.60\pm0.26)$ mJy km s$^{-1}$ arcsec$^{-2}$ for sections 
$b$, $c$ and $d$, with $\Delta V_{\rm FWHM}=20, 25$ and $39$ km s$^{-1}$. CO in $d$ is
offset $20.4\pm2.6$ km s$^{-1}$ from the origin and is responsible for the line asymmetry 
in section II. The other sections yield $\approx0.80-1.56$ mJy km s$^{-1}$ arcsec$^{-2}$ 
($3\sigma$) upper limits. For sections $b, c  ~\& ~d$ we find $\Sigma^{``b"}_{\rm H_2}=1.5$ 
M$_{\odot}$ pc$^{-2}$ ($0.3-5.6$ M$_{\odot}$ pc$^{-2}$), $\Sigma^{``c"}_{\rm H_2}=3.7$ 
M$_{\odot}$ pc$^{-2}$ ($0.5-13.1$ M$_{\odot}$ pc$^{-2}$), and $\Sigma^{``d"}_{\rm H_2}=4.6$ 
M$_{\odot}$ pc$^{-2}$ ($0.7-16.3$ M$_{\odot}$ pc$^{-2}$). Given the sections' corresponding 
$\Sigma_{\rm HI}$ ($28.8$, $36.9~\&~17.7$ M$_{\odot}$ pc$^{-2}$) we derive $f_{\rm mol}=0.05-0.21$.
If we use the largest $\alpha_{\rm CO}$ consistent with the outer ring's metallicity
\citep[$90$ M$_{\odot}$  (K km s$^{-1}$ pc$^{2}$)$^{-1}$;][]{magdis},
$f_{\rm mol}$ only increases to $0.22-0.52$. Nowhere does the Cartwheel's starburst
outer ring appear H$_2$-dominated.

\subsection{H$_2$ Depletion Timescale and Star Formation Law}
The Cartwheel's H$_2$ depletion timescale ($\tau_{\rm mol}=M_{\rm H_2}/SFR$) exhibits
a wide range. In the nucleus and inner ring $\tau_{\rm mol}\approx5-6$ 
Gyr, similar to values derived in local spiral disks (e.g., B08). Much
smaller values characterize the outer ring particularly where $\Sigma_{\rm SFR}$ is high:
$\tau_{\rm mol} = 52$ and $<50$ Myr in $d~\&~e$, which is shorter than the outer ring's
rotational period \citep[$\approx0.5$ Gyr;][]{wof2} and comparable to the main-sequence
lifetime of a B-star. $\tau_{\rm mol}$ is larger in $b$ and $c$ ($640~\& ~189$ Myr) 
and similar to disk averaged $\tau_{\rm mol}$ for starburst galaxies 
\citep[e.g., ][]{kennicutt98}. Elsewhere in the outer ring $\tau_{\rm mol}<160-530$ Myr 
($3\sigma$). 

We show the Cartwheel's molecular and atomic SF laws in Fig.~$5$ and compare
them to those derived for nearby galaxies on comparable spatial
scales where $\Sigma_{\rm SFR}$ is also estimated using H$\alpha + 24~\mu$m emission. 
The Cartwheel's two rings present very different distributions: the $11$ inner 
ring HII complexes and nucleus together are consistent with M51's spatially-resolved 
H$_2$ SF law ($\Sigma_{\rm SFR}\propto\Sigma_{\rm H_2}^{1.37}$; K07);
SF in the outer ring appears to be independent of $\Sigma_{\rm H_2}$ in a manner 
similar to AM0644-741's ring (Fig.~$14$ in HHR). This is in marked contrast 
to the azimuthally averaged molecular SF law derived for the CO-bright disks of the 
HERACLES survey \citep[][L13]{heracles} where $\Sigma_{\rm SFR}\propto\Sigma_{\rm H_2}$.
The HI-dominated outer-disks  of spirals are characterized by reduced metallicity and 
dust/gas ratios and should therefore more closely resemble conditions in the Cartwheel's outer ring.
Fig.~$5$ shows that the averaged outer-disk ($>0.5~r_{\rm 25}$) molecular SF law for the HERACLES 
sample still shows a tight linear relation between $\Sigma_{\rm SFR}$ and $\Sigma_{\rm H_2}$ (S11),
which is again quite unlike the outer ring's SF law.  One might argue that points $b$, $c$ and $d$ 
define a steep power law relation ($\Sigma_{\rm SFR}\propto\Sigma_{\rm H_2}^{3.5}$). 
Any such correlation, however, is greatly weakened by the five H$_2$ upper-limits and the 
wide range in $\Sigma_{\rm H_2}$ allowed by the $\alpha_{\rm CO}$ dispersion, which 
together comprise a large systematic uncertainty. 

HI in the Cartwheel's outer ring reaches $\Sigma_{\rm HI} = 19-65$ M$_{\odot}$
pc$^{-2}$ and, like H$_2$, appears independent of $\Sigma_{\rm SFR}$ (Fig.~$5$). This 
resembles the HI ``saturation'' observed in spiral galaxies 
($\approx8$ M$_{\odot}$ pc$^{-2}$; K07, B08, S11), though with significantly higher 
$\Sigma_{\rm HI}$. Because $f_{\rm mol}$ is so small, the outer ring's combined HI$+$H$_2$ SF law 
is, in effect, the HI SF law (Fig.~$5$). For local spirals the exact opposite opposite is true.
A vertical line can be drawn through all but one point (within the uncertainties) for
the Cartwheel's outer ring, leading to the unusual result that SF appears independent of
the local neutral gas surface density in any form: atomic, molecular or combined.

Normal SF laws break down on small spatial scales and after $\ga30$ Myr for 
individual SF regions (e.g., S10). However, given the $\approx36$~kpc$^{2}$ area  
of regions $a-h$ and the evidence for continuous SF ($\S$1) as the rings propagate, 
other factors must be responsible for the Cartwheel's 
(and AM0644-741's) peculiar SF law. It is moreover extremely unlikely that the
entire SF region populations of both outer rings happen to be in the same brief 
($\approx20$ Myr) CO-faint/H$\alpha$-luminous phase. 

Consumption of the ring's ISM due to high $\Sigma_{\rm SFR}$ could deflect points 
to the left in Fig.~$5$. Indeed, \citet[][]{wof2} noted a significant decrease in 
$\Sigma_{\rm HI}$ across sections $d ~\& ~e$, which is visible in Fig.~$4$. However the 
observed $\Sigma_{\rm SFR}$ would require  $\approx3$ Gyr - roughly $7$-times the ring's 
estimated age - to shift points $d$ and $e$ to their observed positions starting
from any of the spiral galaxy SF laws shown. While gas depletion effects may be at work they 
cannot be entirely responsible for the observed SF law unless $\Sigma_{\rm SFR}$ was much 
higher in the past.

\subsection{The Outer Ring's ISM}
CO emission from the Cartwheel's outer ring indicates an extremely H$_2$ poor ISM
even after accounting for its sub-solar metallicity. In trying to understand
this situation we discount the possibility that we are viewing the Cartwheel (and
AM0644-741) just as their molecular reservoirs have been exhausted, since both rings'
high $\Sigma_{\rm HI}$ favors the rapid conversion of HI to H$_2$. 
It may be the case that conditions in the starburst ring lead to
enhanced destruction of H$_2$ so that even a metallicity-appropriate $\alpha_{\rm CO}$
reveals little molecular gas. This however leaves the question of how SF is triggered so
efficiently in the remaining H$_2$. Alternately, the 
starburst rings may act to make CO an unreliable proxy for H$_2$. 
The ability of CO to trace H$_2$ in a molecular cloud (MC) depends
sensitively on the cloud's net visual extinction, $\overline{A}_{\rm V}$, which is 
proportional to the product of its metallicity and H$_2$ column density ($N_{\rm H_2}$) 
\citep[e.g.,][]{wolf2010, clark2015, lee2015}. Small or low-$n_{\rm H}$ (or both) MCs 
in active star forming regions may inadequately shield the fragile
CO molecule and as a result become extremely CO-faint.  Reduced metallicity 
substantially compounds this effect. We propose that the ISM's long confinement 
in the starburst rings of these systems can result in a state characterized by 
smaller MCs and a diffuse H$_2$ component due to the accumulated damage from embedded 
OB stars and SNe (HHR). This helps explain the extremely low average $n_{\rm H}$ 
($1-5$ cm$^{-3}$) we derive for AM0644-741's starburst quadrant.  If this characterizes the overall 
state of the ISM then low-$N_{\rm H_2}$ MCs may dominate high $\Sigma_{\rm SFR}$
regions, resulting in a CO-faint molecular ISM and low inferred $M_{\rm H_2}$. A further dependence 
on metallicity and local $\Sigma_{\rm SFR}$ would be expected to produce a peculiar SF law 
like that in Fig.$~5$.  Conversely, the inner ring's 
$\approx$normal $\Sigma_{\rm H_2}$,  $f_{\rm mol}$, $\tau_{\rm mol}$ and SF law 
would be a consequence of its lower  $\Sigma_{\rm SFR}$ and age ($70$ vs. $440$ Myr): 
the inner ring's ISM has yet to reach the outer ring's state.

These possibilities can be further tested observationally.
Dense ($n>10^{4}$ cm$^{-3}$) gas in cloud cores should be better protected from
the destructive effects of massive stars and is, moreover, the component directly connected 
with SF \citep[][]{solomon92,gaosolomon}. Observations of the Cartwheel and AM0644-741
using HCN or HCO+ rotational transitions may uncover $\approx$normal SF laws. 
Infrared/submillimeter dust emission would additionally provide reliable estimates 
of H$_2$ in the rings independent of CO \citep[e.g., ][]{leroy09}. Our analysis further
suggests that high $\Sigma_{\rm SFR}$ and long residence time are both required 
to reach the peculiar ISM state we infer in starburst rings. Observations of 
{\em young} ring galaxies (i.e., age $\lesssim100$ Myr) would in this case be expected to 
show more normal $f_{\rm mol}$ and SF laws. Preliminary support for this is suggested by 
the global $f_{\rm mol}$ ($0.40$) of the young ($age\approx35$~Myr) ring galaxy Arp$~147$. 
State of the art ring galaxy models incorporating a multi-phase ISM and the effects 
of massive stars and SNe will also help place these ideas on a firmer physical basis.

\section{Conclusions}

Both of the Cartwheel's rings are gas rich, expanding and forming stars, 
though with large differences in their relative SFR, $\Sigma_{\rm SFR}$ 
and ages. They appear dominated by different ISM phases: molecular in the 
inner and atomic in the outer.  Even with a metallicity appropriate $\alpha_{\rm CO}$ 
the outer ring's H$_2$ content appears strikingly low for its star formation
activity, which leads directly to its unusually low $\tau_{\rm mol}$ and peculiar
SF law. By contrast  the inner ring appears normal. Similar results were found in 
AM0644-741's $42$ kpc diameter starburst ring, leading us to reject the possibility that we are
observing both ring galaxies just as their H$_2$ reservoirs are consumed or
when their SF regions simultaneously attain a brief CO-faint/H$\alpha$-luminous phase. 
Either highly efficient SF occurs in a H$_2$-poor ISM in the rings of these systems or 
rotational transitions of $^{12}$CO do not accurately trace H$_2$. Our observations 
are consistent with the second explanation if the ISM of 
both starburst rings possess abundant small and/or low-$N_{\rm H_2}$ MCs that 
inadequately shield CO molecules. This is a direct consequence of the ISM's long residence in the 
outer rings and is very likely compounded by sub-solar metallicity in addition to (limited) gas depletion 
effects. If so, the outer ring's $M_{\rm H_2}$ and 
$\Sigma_{\rm H_2}$ may be substantially greater than that inferred by CO.
We expect such a state to characterize large and robustly star forming ring galaxies
and possibly starburst nuclear rings associated with gravitational resonances.
The large departures from a standard SF law we derive
in the outer rings of the Cartwheel and AM0644-741 represent interesting puzzles
that a comprehensive theory of star formation should be able to address.

\acknowledgments
This Letter makes use of ALMA dataset ALMA$\#$2012.1.00720.S.
ALMA is a partnership of ESO (representing its member states), NSF (USA), and NINS
(Japan), together with NRC (Canada) and NSC and ASIAA (Taiwan), in cooperation with the 
Republic of Chile. The Joint ALMA Observatory is operated by ESO, AUI/NRAO, and NAOJ.
Based on observations collected at the European Organization for Astronomical Research
in the Southern Hemisphere under ESO programs 66.B-0666(A) and 66.B-0666(B).
The National Radio Astronomy Observatory is a facility of the National Science Foundation 
operated under cooperative agreement by Associated Universities, Inc. This work is based
in part on observations made with the {\em Spitzer Space Telescope}, which is operated by the 
Jet Propulsion Laboratory, California Institute of Technology under a contract with NASA.
The authors thank K. Sheth for the initial reduction of the ALMA data, G. Cecil 
for his role in the spectroscopic observations, and the Paranal observatory staff for 
the VLT service observations. We thank an anonymous referee for constructive comments.

\begin{figure}
\figurenum{1}
\epsscale{1.0}
\plotone{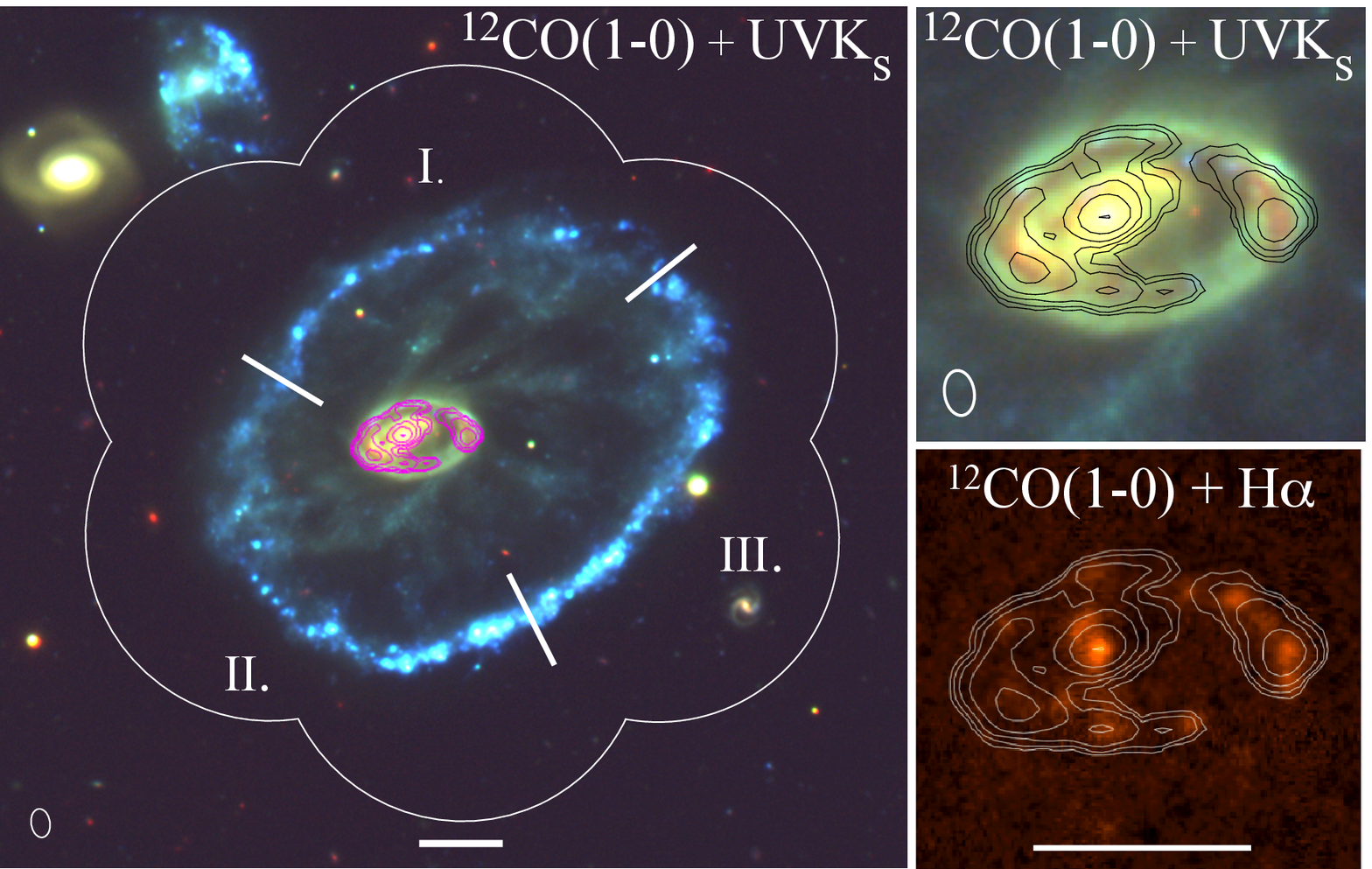}
\caption{Molecular gas in the Cartwheel. (left) {\it natural}-weighted $^{12}$CO(J=1-0) 
emission (moment-$0$) contoured on a VLT UVKs three-color image with six of the 
seven-position mosaic primary beams ($53\arcsec$ FWHM) 
outlined. Contours correspond to $\Sigma_{\rm H_2} = 6.8, 15.9, 37.3, 87.2 ~\& ~204.0$ 
M$_{\odot}$ pc$^{-2}$ for a Galactic $\alpha_{\rm CO}$ and include helium. Only the 
inner ring and nucleus are directly detected. The synthesized beam
($2.43\arcsec\times1.50\arcsec$) is shown at bottom-left.  
(top-right) A close-up of the inner ring's CO emission, including H$\alpha$ emission 
(bottom-right). Both scale-bars are $10\arcsec$ ($6.1$~kpc).}
\end{figure}

\begin{figure}
\figurenum{2}
\epsscale{0.5}
\plotone{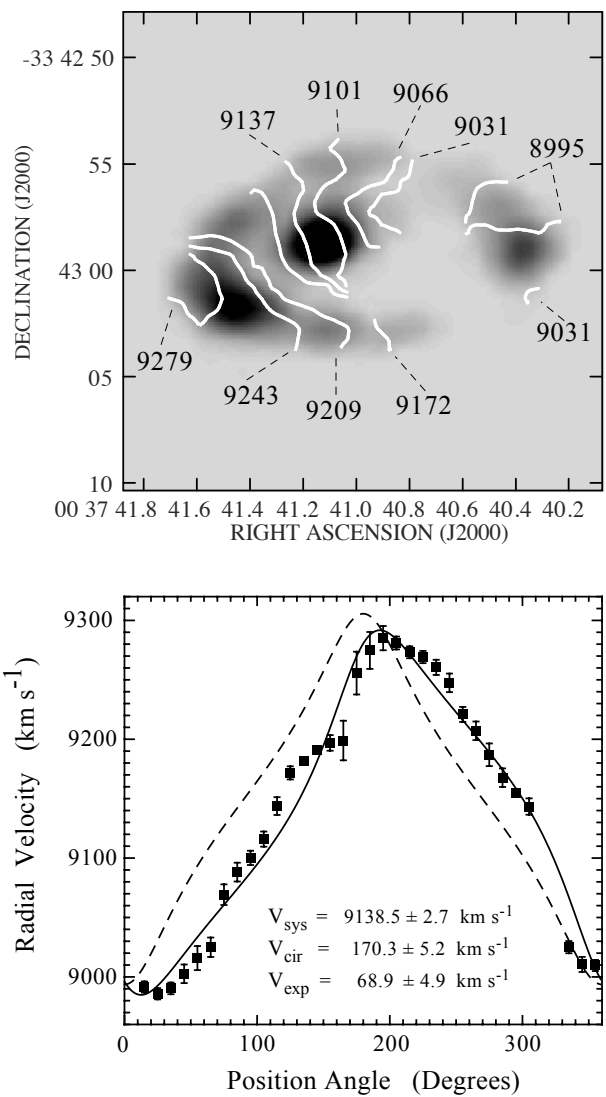}
\caption{The inner ring's CO kinematics. (top) Isovelocity contours (optical/heliocentric)
from the moment-$1$ map shown plotted on the moment-$0$ greyscale image. (bottom) The inner 
ring's radial velocity-position angle diagram showing least-square fits for inclined 
rotating circular rings with (solid) and without (dashed) expansion.}
\end{figure}

\begin{figure}
\figurenum{3}
\epsscale{0.3}
\plotone{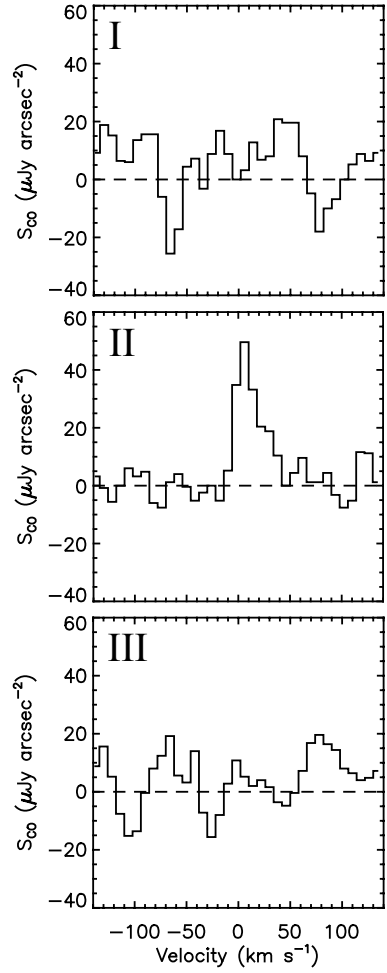}
\caption{Stack averaged $^{12}$CO(J=1-0) spectra for sections I, II and III of
the Cartwheel's outer ring as defined in Fig.~$1$. Molecular gas is detected 
only in section II.}
\end{figure}

\begin{figure}
\figurenum{4}
\epsscale{0.8}
\plotone{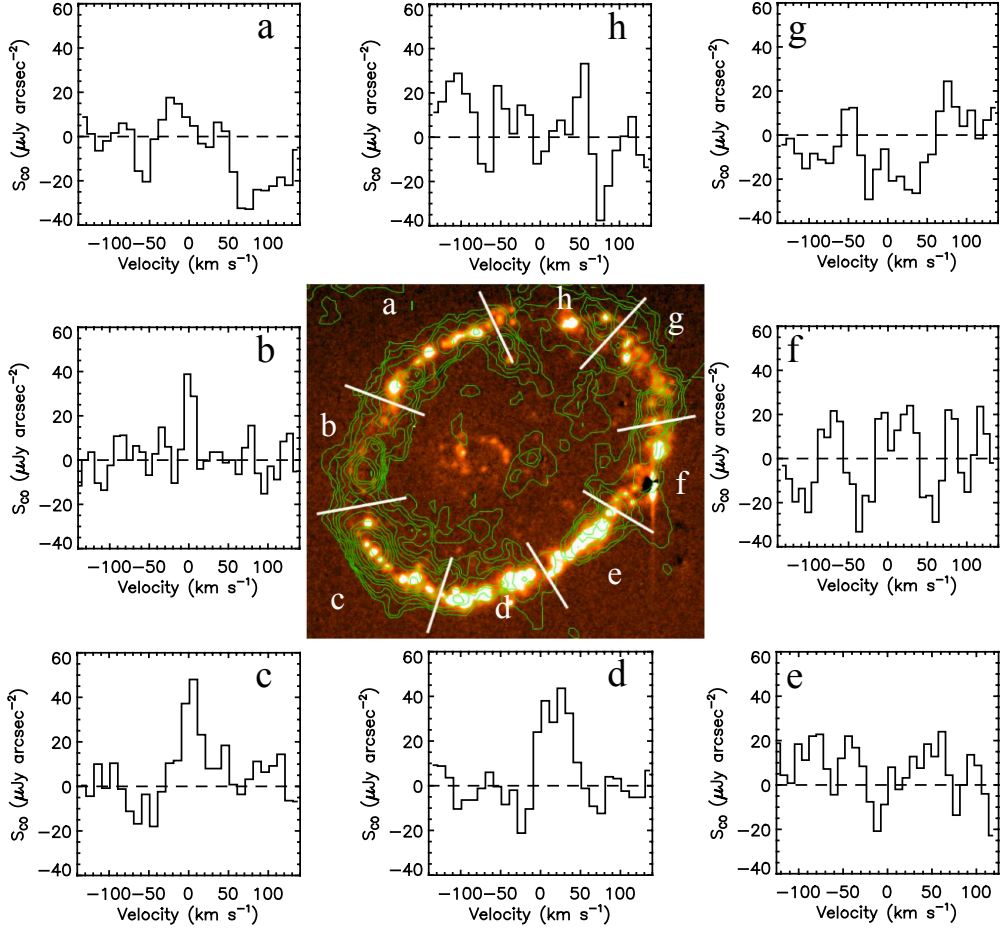}
\caption{Stack averaged $^{12}$CO(J=1-0) spectra for the eight sections of the Cartwheel's 
outer ring as defined for the H$\alpha$ image (color in the electronic version) at center. 
Molecular gas is detected only at $b-d$, which corresponds to II in Fig.~$1$. Contours 
represent $\Sigma_{\rm HI} = 2.7, 4.1, 6.3, 9.4, 14.3, 21.5 ~\& ~32.6$ M$_{\odot}$ 
pc$^{-2}$ and include helium.\citep[][]{wof2}.}
\end{figure}

\begin{figure}
\figurenum{5}
\epsscale{.9}
\plotone{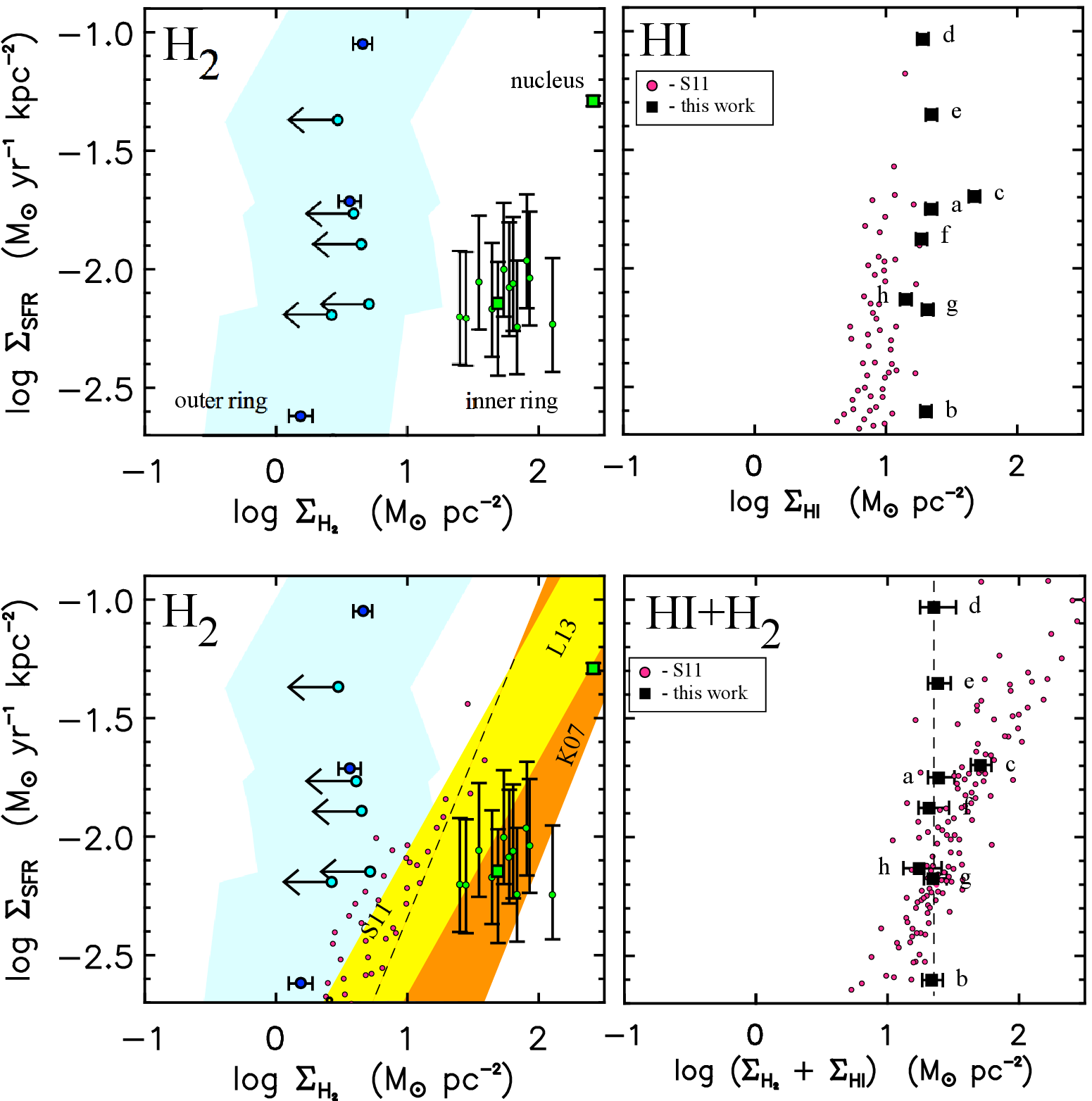}
\caption{The Cartwheel's SF law. (top-left) Derived H$_2$ SF law for the outer
and inner rings. The nucleus and averaged inner ring are represented by squares. The shaded area 
indicates the allowed $\Sigma_{\rm H_2}$ given $\alpha_{\rm CO}$'s dispersion in 
\citet[][]{magdis}. (bottom-left) The Cartwheel's molecular SF law compared with M51's spatially-resolved 
molecular SF law (K07), the azimuthally averaged SF law for the HERACLES sample of local spirals and irregulars (L13),
and their outer ($>0.5~r_{\rm 25}$) HI-dominated disks only (S11). The latter assumes a Galactic $\alpha_{\rm CO}$. 
 (top-right) The Cartwheel's HI SF law 
compared to the HI-dominated outer disks in the HERACLES survey (S11). (bottom-right) The outer ring's total gas SF 
law. $\Sigma_{\rm SFR}$ appears independent of $\Sigma_{\rm HI + H_2}$. The corresponding 
total SF law derived with the HERACLES sample is also shown.}
\end{figure}


\begin{thebibliography}{}

\bibitem[Appleton et al.(1996)]{appleton1996}
   Appleton, P. N., Struck-Marcell, C., Bransford, M. A., Charmandaris, V., 
   Marston, A. et al. 1996, in I.A.U. Symp. 171, New Light on Galaxy Evolution,  
   ed. R. Bender \& R. L. Davies (Kluwer, Dordrecht), 337

\bibitem[Bigiel et al.(2008)]{bigiel2008}
   Bigiel, F., Leroy, A., Walter, F., Brinks, E., de Blok, W. et al., 2008, \aj, 136, 2846 (B08)

\bibitem[Clark \& Glover(2015)]{clark2015}
   Clark, P. C., \& Glover, S. C. 2015, \mnras, 452, 2057

\bibitem[Fosbury \& Hawarden(1977)]{fh77}
   Fosbury, R. A. E., \& Hawarden, T. G. 1977, \mnras, 178, 473

\bibitem[Gao \& Solomon(2004)]{gaosolomon}
   Gao, Y. \& Solomon, P. 2004, \apj, 606, 271

\bibitem[Higdon(1995)]{wof1}
   Higdon, J. L. 1995, \apj, 455, 524

\bibitem[Higdon(1996)]{wof2}
   Higdon, J. L. 1996, \apj, 467, 241

\bibitem[Higdon, Higdon \& Rand(2011)]{wof4}
   Higdon, J. L., Higdon, S. J. U., \& Rand, R. J. 2011, \apj, 739, 97  (HHR)

\bibitem[Horellou et al.(1998)]{horellou}
   Horellou, C., Charmandarys, V., Combes, F., et al. 1998, \aap, 340, 51

\bibitem[Horellou et al.(1995)]{horellou95}
   Horellou, C., Casoli, F., Combes, F. et al., 1995, \aap, 298, 743

\bibitem[Jogee et al.(2009)]{jogee09}
   Jogee, S., Miller, S. H., Penner, K., et al. 2009, \apj, 697, 1971

\bibitem[Kennicutt(1998)]{kennicutt98}
   Kennicutt, R. C. 1998, \apj, 498, 541

\bibitem[Kennicutt et al.(2007)]{kennicutt2007} 
   Kennicutt, R. C., Calzetti, D., Walter, F., et al. 2007, \apj, 671, 333  (K07)

\bibitem[Lee et al.(2015)]{lee2015}
   Lee, C., Leroy, A. K., Schnee, S., et al. 2015, \mnras, 450, 2708

\bibitem[Leroy et al.(2009a)]{heracles}
   Leroy, A. K., Walter, F., Bigiel, F. et al. 2009, \aj, 137, 4670

\bibitem[Leroy et al.(2009b)]{leroy09}
   Leroy, A. K., Bolatto, A., Bot, C., et al. 2009, \apj, 702, 352 

\bibitem[Leroy et al.(2013)]{leroy2013}
   Leroy, A., Walter, F., Sandstrom, K., Schruba, A., Munoz-Mateos, J. et al. 2013, \aj, 146, 19 (L13)

\bibitem[Lynds \& Toomre(1977)]{lt77}
   Lynds, R. \& Toomre, A. 1977, \apj, 209, 382

\bibitem[Magdis et al.(2011)]{magdis}
   Magdis, G. E., Daddi, E., Elbaz, D., Sargent, M., Dickinson, M. et al., 2011, \apj, 740, 15

\bibitem[Mapelli et al.(2008)]{mapelli2008}
   Mapelli, M., Moore, B., Ripamonti, E., et al. 2008, \mnras, 383, 1223

\bibitem[Mapelli \& Mayer(2012)]{mapelli_mayer2012}
   Mapelli, M., \& Mayer, L. 2012, \mnras, 420, 1158

\bibitem[Marcum, Appleton \& Higdon(1993)]{marcum93}
   Marcum, P. M., Appleton, P., \& Higdon, J. L. 1993, \apj, 399, 57

\bibitem[McMullin et al.(2007)]{mcmullin}
   McMullin, J. P., Waters, B., Schiebel, D., Young, W., \& Golap, K. 2007, in ASP
Conf. Ser. 376, Astronomical Data Analysis Software and Systems XVI, ed.
R. A. Shaw, F. Hill, \& D. J. Bell (San Francisco, CA: ASP), 127

\bibitem[Mihos \& Hernquist(1994)]{mihos_hernquist}
   Mihos, J. C., \& Hernquist, L. 1994, \apj, 437, 611

\bibitem[Nagao et al.(2006)]{nagao}
   Nagao, T., Maiolino, R., \& Marconi, A. \ 2006, \aap, 459, 85

\bibitem[Romano, Mayya \& Vorobyov(2008)]{romano2008}
   Romano, R., Mayya, Y., \& Vorobyov, E. 2008, \aj, 136, 1259

\bibitem[Schruba et al.(2011)]{schruba2011}
  Schruba, A., Leroy, A., Walter, F., Bigiel, F., Brinks, E. et al. 2011, \aj, 142, 37 (S11)

\bibitem[Smail et al.(1997)]{smail97}
   Smail, I., Ivison, R. J., \& Blain, A. W., 1997, \apj, 490, L5

\bibitem[Solomon et al.(1992)]{solomon92} 
   Solomon, P. M., Downes, D., \& Radford, S. J., 1992, \apj, 387, 55 

\bibitem[Solomon \& Vanden Bout(2005)]{solomonvdb05}
   Solomon, P. M., \& Vanden Bout, P. A. 2005, ARAA, 43, 677

\bibitem[Strong et al.(1988)]{strong}
   Strong, A. W., Bloemen, J., Dame, T., et al. 1988, \aap, 207, 1

\bibitem[Struck \& Higdon(1993)]{struckhigdon}
   Struck, C. J., \& Higdon, J. L. 1993, \apj, 411, 108

\bibitem[Tacconi et al.(2008)]{tacconi08}
   Tacconi, L. J., Genzel, R., Smail, I., et al. 2008, \apj, 680, 246

\bibitem[White \& Rees(1978)]{whiterees78}
   White, S. D., \& Rees, M. J. 1978, \mnras, 183, 341

\bibitem[Wolfire, Hollenbach \& McKee(2010)]{wolf2010}
   Wolfire, M. G., Hollenbach, D., \& McKee, C. F. 2010, \apj, 716, 1191

\end{thebibliography}
\end{document}